\documentclass[a4paper,11pt]{article}
\usepackage{amsfonts,amssymb,graphicx,amsmath}

\newtheorem{theorem}{Theorem} [section]

\newtheorem{corollary}[theorem]{Corollary}

\newtheorem{definition}[theorem]{Definition}
\newtheorem{example}[theorem]{Example}

\newtheorem{lemma}[theorem]{Lemma}

\newenvironment{proof}[1][Proof]{\noindent\textbf{#1.} }{\ \rule{0.5em}{0.5em}}
\begin{document}

\author{Luis A. Guardiola\thanks{%
Operations Research Center, Universidad Miguel Hern\'{a}ndez, Edificio
Torretamarit, Avda. de la Universidad s.n, 03202 Elche (Alicante), Spain.}
\thanks{%
Corresponding author.},\ Ana Meca$^{\dag }$ and Judith Timmer\thanks{%
Stochastic Operations Research Group, Department of Applied Mathematics,
University of Twente, Enschede, The Netherlands.}}
\title{Cooperation and profit allocation%
%TCIMACRO{\TeXButton{TeX field}{\\ } }%
%BeginExpansion
\\
%EndExpansion
in distribution chains\thanks{ This work was partially supported
by the Spanish Ministry of Education and Science and Generalitat
Valenciana (grants MTM2005-09184-C02-02, CSD2006-00032,
ACOMP06/040). The authors acknowledge valuable comments made by
two anonymous referees. }}
\date{}
\maketitle

\begin{abstract}
We study the coordination of actions and the allocation of profit in supply
chains under decentralized control in which a single supplier supplies
several retailers with goods for replenishment of stocks. The goal of the
supplier and the retailers is to maximize their individual profits. Since
the outcome under decentralized control is inefficient, cooperation among
firms by means of coordination of actions may improve the individual
profits. Cooperation is studied by means of cooperative game theory. Among
others we show that the corresponding games are balanced and we propose a
stable solution concept for these games.


\end{abstract}

\vfill

\section{Introduction}

The study of the coordination of actions and the allocation of profits in
distribution chains are becoming popular topics nowadays. In the literature
on distribution channel coordination both cooperative and noncooperative
game theory are used quite often to investigate these topics. There are some
differences between the use of noncooperative and cooperative game theory.
When applying noncooperative game theory it is assumed that each company in
the supply chain is an independent decision-maker that acts in accordance
with its objective. One of the main points of concern is whether some
proposed coordination mechanism coordinates the supply chain, that is
maximizes the total profit of the supply chain, under a competitive
framework. In contrast, cooperative game theory assumes that companies can
make binding agreements. One of the main questions here is whether the
cooperation is stable, that is, whether there are allocations of the total
profit among the companies such that no group of companies would like to
cooperate on its own.

Channel coordination is an issue that has been investigated under several
perspectives. Its main goal is to improve total profits and welfare. In this
paper, we focus on the analysis of supply chains by means of cooperative
games. In particular, we consider single period distribution chains with a
single product. In such a supply chain, retailers place one-time orders for
the product at the supplier. After production, the supplier delivers the
goods to the retailers via a warehouse. This warehouse acts as an
intermediary without costs or revenues. When the goods arrive, the
noncompeting retailers sell these on their own separate markets. The larger
the quantity that is put on the market, the lower the expected revenue per
unit for the retailer. Each retailer chooses its order quantity such that
its profit is maximized.

The retailer pays the supplier a wholesale price per unit product ordered
and delivered. This price is a decreasing function of the quantity ordered.
Hence, incentives for cooperation among retailers exist. If the retailers
combine their orders into one large order then they enjoy a lower wholesale
price per unit. They can do so because the warehouse only informs the
supplier about the quantities ordered and not about which retailer orders
how much. Besides, retailers may want to cooperate with the supplier which
implies a further reduction in cost inefficiency due to the absence of the
intermediate wholesale prices. Obviously, the total profit under full
cooperation is larger than the sum of the individual profits.

Because of the incentives for cooperation, we use cooperative game theory to
study these distribution chains. For each chain we define a corresponding
cooperative game in which the supplier and the retailers are the players.
The value of a coalition of players equals the optimal joint profit they can
achieve. We show that the core of such a game is never empty, that is, all
companies in the chain are willing to cooperate because there exist stable
distributions of the total profit among the companies upon which no
coalition can improve. Any distribution of profits that belongs to the core
has a nice interpretation in terms of the underlying distribution chain.
Further, we introduce a specific allocation of the total profit for
distribution chains, the so-called minimal-gain-per-capita (mgpc) solution.
This solution is a stable distribution of the profits, that is, it always
belongs to the core of the game, and it possess several nice properties. In
particular, it takes into account the importance of the supplier to achieve
full cooperation. Finally, a characterization of the mgpc-solution is
provided.

Our paper contributes to the emerging literature on the analysis
of problems in Operations Research by means of cooperative game
theory. Some recent papers in this area are \cite{GuMe04b},
\cite{Mar05}, \cite{GuMe06}, \cite{GuMe07} and \cite{VdBo07}, and
for a survey we refer to \cite{BoHa01}. Closely related to our
work are papers that focus on cooperation in supply chains by
means of cooperative games. Perhaps one of the first to do so is
\cite{Robi93}, who uses cooperative game theory to study the
allocation of joint inventory control costs among multiple
retailers. The author shows that the Shapley value \cite{SH53} of
the corresponding cooperative game belongs to the core, and as
such is a fair and suitable allocation for sharing the costs. In
\cite{WaPa94} a three-player newsvendor game is analyzed by means
of both noncooperative and cooperative game theory. Among others,
conditions are given such that the core of related cooperative
games is nonempty and the authors conclude that full cooperation
cannot always be obtained. \cite{SlFr01} studies joint ordering by
multiple retailers. The authors study a related cooperative game
in which the retailers are the players. Their main result is that
this game
has a nonempty core. This result is extended in subsequent studies \cite%
{OzFr04,SlFr05}. Noncooperative game theory is also used to study problems
in supply chains namely, among others, to study coordination mechanisms,
like contracts, under horizontal and vertical competition. We refer to \cite%
{BeFe03, BeFe05, CaLa01, CaLa05, KriKa01, Ta92, TsNa98, RyMa00}
for reviews on analyses of contracts. Finally, \cite{Mi05}
analyses collaboration for the economic order quantity model by
using bargaining concepts from non-cooperative and cooperative
game theory.

The contribution of our paper to the literature is twofold. First, we
include the supplier in our analysis and study cooperation among retailers
and the supplier. Second, we introduce a tailor-made allocation of the joint
profit that always belongs to the core of the game. The above-mentioned
literature only considers cooperation among retailers and hardly pays
attention to suitable allocations of the joint benefits.

The contents of this paper are as follows. In the next section we introduce
the necessary concepts of cooperative game theory. Our model of a single
period distribution chain is introduced and studied in section \ref%
{SecRSProblems}. A related cooperative game, the RS-game, is studied in
section \ref{SecRSGames}. Thereafter, in section \ref{SecBalancedness}, we
show that RS-games always have a nonempty core. Any allocation in the core
is shown to have a nice interpretation in terms of its underlying
distribution chain. In section \ref{SecSolution} we introduce and study the
mgpc-solution for RS-games. We show that it belongs to the core and
characterize it. Section \ref{SecConcl} concludes.

\section{Preliminaries cooperative game theory}

A cooperative (benefit) game with transferable utility (TU game) is a pair $%
(N,v)$ where $N=\left\{ 1,2,...,n\right\} $ is the finite player set. Let $%
P(N)$ be the set containing all subsets of $N$ then $v:P(N)\rightarrow
\mathbb{R}$ is the characteristic function of the game satisfying $%
v(\emptyset )=0.$ A coalition is a nonempty subset of $N$. The subgame $%
v_{S} $ related to coalition $S$ is the restriction of the mapping $v$ to
the subcoalitions of $S.$ We denote by $s$ the cardinality of the set $%
S\subseteq N$, i.e. $card(S)=s.$ A benefit vector, or allocation, is denoted
by $x\in \mathbb{R}^{n}$. The core of the game $(N,v)$ consists of those
allocations of $v(N)$ in which each coalition receives at least its benefit
as prescribed by the characteristic function: $Core(N,v)=\left\{ x\in
\mathbb{R}^{n}\left/ \sum_{i\in N}x_{i}=v(N)\text{ and }\sum_{i\in
S}x_{i}\geq v(S)\text{ for all }S\subset N\right. \right\} .$ A
core-allocation $x\in Core(N,v)$ is both efficient, that is $\sum_{i\in
N}x_{i}=v(N)$, and it satisfies the coalitional stability property, that is $%
\sum_{i\in S}x_{i}\geq v(S)$ for all $S\subset N$. A game $(N,v)$ is
balanced if and only if it has a nonempty core \cite{B63,SH67}. It is a
totally balanced game if all its subgames are balanced.

A game $(N,v)$ is strict monotone increasing if $v(S)<v(T)$ for all $%
S\subset T.$ It is superadditive if $v(S\cup T)\geq $ $v(S)+v(T)$ holds for
all disjoint coalitions $S$ and $T$. In a superadditive game, it is always
beneficial for two disjoint coalitions to cooperate and form a larger
coalition. A well-known class of balanced and superadditive games is the
class of convex games \cite{SH71}$.$ A game $(N,v)$ is convex if $v(S\cup
\{i\})-v(S)\leq v(T\cup \{i\})-v(T)$ for all $i\in N$ and for all $%
S\subseteq T\subseteq N\setminus \{i\}$.

A single-valued solution $\varphi $\ for TU games $(N,v)$ is a map $\varphi
:\Gamma ^{N}\rightarrow \mathbb{R}^{N}$\ where $\Gamma ^{N}$\ is the class
of TU-games with player set $N$. The payoff to player $i\in N$ in game $v\in
\Gamma ^{N}$ according to this solution is denoted by $\varphi _{i}(v)$\ and
$\varphi (v)=\left( \varphi _{i}(v)\right) _{i\in N}$.
%One example of such a solution is the Shapley value $Sh(v)$ \cite{SH53}. For a game $(N,v)$ it is defined by%
%\begin{equation*}
%Sh_{i}(v)=\sum\limits_{S\subseteq N\backslash \{i\}}\frac{s!(n-s-1)!}{n!}%
%\left[ v(S\cup \{i\})-v(S)\right]
%\end{equation*}%
%for all $i\in N$.

We denote by $\mathbb{R}_{+}^{n}$\ the set of $n$-dimensional real vectors
whose components are nonnegative. For $a,b\in \mathbb{R}$, $a<b$,\ $[a,b]$\
is a closed interval and $(a,b)$ an open interval in $\mathbb{R}$.

\section{Retailer-Supplier problems\label{SecRSProblems}}

In this paper we study single period models of distribution chains involving
a single product. In these chains, a supplier replenishes his goods to
several retailers via a warehouse. One can think, for example, of a car
manufacturer (supplier) who is about to produce a car with special features
that will only be temporarily available. This is a single period model in
which the local car dealers (retailers) have one opportunity to place their
orders for this special car at the national importer (warehouse), who passes
the national orders to the car manufacturer. In this section we first
concentrate on a chain with a single retailer. Chains with multiple
retailers are considered in the next section.

The retailer places a one-time order, say of size $q$ units, for the good at
the warehouse, who passes this information to the supplier. The costs of
this order are $w(q)$ per unit for the retailer, where the wholesale price
function $w:\mathbb{R}_{+}\rightarrow (c,+\infty )$\ (i.e. $w(q)>c$\ for all
$q\geq 0$) is a decreasing and continuous function. It's decreasing nature
represents quantity discounts provided by the supplier: the more the
retailer orders, the lower the price per unit he has to pay. The ordered
goods are produced by the supplier at a cost of $c$ per unit. After
production the goods are shipped to the warehouse, who acts as an
intermediate party with no costs or benefits. The warehouse sends the goods
to the retailer, who in turn sells the goods on the market. The expected
revenue of the retailer is $p(q)$ per unit of the good, given the supply of $%
q$ units on the consumer market. The expected price function $p:\mathbb{R}%
_{+}\rightarrow \mathbb{R}$\ is decreasing and continuous in $q$, satisfies $%
p(0)>w(0)$\ and there exists a quantity $q>0$\ such that $p(q)=c$. %
Notice that the latter condition makes sense from an economic
point of view since if $p(q)>c$\ for all $q$\ then the parties in
the chain can obtain arbitrary large profits.

A retailer-supplier problem\ (henceforth: RS-problem) is denoted by the
tuple $(c,w,p)$. Given such a problem let $\mathbb{Q}=\{q\in \mathbb{R}%
_{+}\mid p(q)\geq w(q)\}$\ be the set of feasible order sizes, that is,
those order sizes that result in a nonnegative profit margin for the
retailer. Notice that $\mathbb{Q\neq \emptyset }$ because $p(0)>w(0)$.

The retailer determines his order quantity $q$ such that his (expected)
profit is maximized:%
\begin{eqnarray*}
\max &&\Pi ^{ret}\left( q;w\left( q\right) \right) =\left( p(q)-w(q)\right) q
\\
\mbox{s.t.} &&q\in \mathbb{Q}.
\end{eqnarray*}%
This optimization problem always has an optimal solution/order size $q$
since $\mathbb{Q}$ is a compact set and $\Pi ^{ret}\left( q;w\left( q\right)
\right) $ is continuous on $\mathbb{Q}$. Given the retailer's order size $q$%
, the supplier's profit equals%
\begin{equation*}
\Pi ^{sup}\left( q;w\left( q\right) \right) =\left( w(q)-c\right) q.
\end{equation*}%
The two examples\footnote{%
The reader should notice that some calculations may be needed to check the
examples.} below show that the retailer may have one or more optimal order
sizes.

\begin{example}
\label{Ex_f1}Let $(c,w,p)$ be an RS-problem with $c=2,$ $p(q)=7-q$ and
\begin{equation*}
w(q)=\left\{
\begin{array}{ll}
5, & 0\leq q\leq 1, \\
2+3/q, & q>1.%
\end{array}%
\right.
\end{equation*}%
The retailer solves%
\begin{eqnarray*}
\max &&\Pi ^{ret}\left( q;w\left( q\right) \right) \\
\mbox{s.t.} &&q\in \mathbb{Q}=\left[
0,\frac{5+\sqrt{13}}{2}\right]
\end{eqnarray*}%
where%
\begin{equation*}
\Pi ^{ret}\left( q;w\left( q\right) \right) =\left\{
\begin{array}{ll}
2q-q^{2}, & 0\leq q\leq 1 \\
-q^{2}+5q-3, & q>1.%
\end{array}%
\right.
\end{equation*}%
The unique optimal order size is $q^{\ast }=2\frac{1}{2}$, leading to a
profit of $\Pi ^{ret}\left( q^{\ast };w\left( q^{\ast }\right) \right) =3%
\frac{1}{4}$ for the retailer and $\Pi ^{sup}\left( q^{\ast };w\left(
q^{\ast }\right) \right) =3$ for the supplier.
\end{example}

\begin{example}
\label{Ex_f2}Let $(c,w,p)$ be an RS-problem with $c=1,$%
\begin{equation*}
p(q)=\left\{
\begin{array}{ll}
5, & 0\leq q\leq 1, \\
6-q, & 1<q\leq 2, \\
5-\frac{q}{2}, & q>2,%
\end{array}%
\right.
\end{equation*}%
and%
\begin{equation*}
w(q)=\left\{
\begin{array}{ll}
4, & 0\leq q\leq 1, \\
3+\frac{1}{q}, & 1<q\leq 2, \\
2\frac{1}{4}+\frac{5}{2q}, & 2<q\leq 2\frac{1}{2}, \\
3\frac{1}{4}, & q>2\frac{1}{2}.%
\end{array}%
\right.
\end{equation*}%
Now the retailer solves%
\begin{eqnarray*}
\max  &&\Pi ^{ret}\left( q;w\left( q\right) \right)  \\
\mbox{s.t.} &&q\in \mathbb{Q}=\left[ 0,3\frac{1}{2}\right]
\end{eqnarray*}%
where%
\begin{equation*}
\Pi ^{ret}\left( q;w\left( q\right) \right) =\left\{
\begin{array}{ll}
q, & 0\leq q\leq 1, \\
-q^{2}+3q-1, & 1<q\leq 2, \\
-\frac{1}{2}q^{2}+2\frac{3}{4}q-2\frac{1}{2}, & 2<q\leq 2\frac{1}{2}, \\
-\frac{1}{2}q^{2}+1\frac{3}{4}q, & q>2\frac{1}{2}.%
\end{array}%
\right.
\end{equation*}%
This leads to two optimal order sizes, namely $q_{a}^{\ast }=1\frac{1}{2}$
and $q_{b}^{\ast }=2\frac{1}{2}.$ In either case the optimal benefit for the
retailer is $1\frac{1}{4}$. The profit for the supplier is either $\Pi
^{sup}\left( q_{a}^{\ast };w\left( q_{a}^{\ast }\right) \right) =4$ or $\Pi
^{sup}\left( q_{b}^{\ast };w\left( q_{b}^{\ast }\right) \right) =5\frac{5}{8}
$. The reader may notice that the supplier prefers $q_{b}^{\ast }$%
\ over $q_{a}^{\ast }$\ because it results in larger
profits. However, he has no means to induce it. The only way the
supplier would be able to influence the retailer's decision is by
changing the
wholesale price $w(q)$, but that is outside the scope of this paper.%

\end{example}

\section{Retailers-Supplier games\label{SecRSGames}}

In this section we address a natural extension of the RS-problem, namely, we
consider single period models of distribution chains with a supplier, a
warehouse and multiple retailers. Each of the retailers places its order at
the warehouse, who passes the order sizes to the supplier. The retailers
have the possibility to cooperate among each other and place a joint order,
which results in a lower wholesale price per unit. The supplier will not
know about the cooperation since he only receives the order sizes and does
not know which order size belongs to which retailer(s). Hence the presence
of the warehouse allows the retailers to save money by cooperation.
Furthermore, a group of retailers may cooperate with the supplier via the
warehouse. In this case, the warehouse provides all parties with the
necessary information to achieve cooperation. After delivery of the goods
from the supplier via the warehouse to the retailers, each retailer sells
its goods on its local consumer market. These markets are independent from
one another, implying that the retailers do not compete for customers.

Let $N=\{1,...,n\}$ be the set of retailers and denote the supplier by $0$.
Then $N_{0}=N\cup \{0\}$ is the set of all companies in the chain.
Similarly, we define $S_{0}=S\cup \{0\}$\ for all $S\subseteq N$. A
retailers-supplier situation (RS-situation) is a tuple $(N_{0},c,w,P)$ with $%
P=(p_{1},...,p_{n})$, $p_{i}$ is the expected price function of retailer $i$%
, and the tuple $(c,w,p_{i})$ is an RS-problem for any retailer $i$.

In such an RS-situation two types of cooperation among the companies may
occur, namely cooperation excluding or including the supplier. First,
cooperation among retailers is profitable since the firms may place one
large joint order for the good and thus enjoy a quantity discount provided
by the supplier. If $q_{S}=\sum_{i\in S}q_{i}$ denotes the total order size
by a coalition $S\subseteq N$ of retailers then the joint benefit of this
coalition equals%
\begin{eqnarray*}
\max &&\sum_{i\in S}\left( p_{i}(q_{i})-w\left( q_{S}\right) \right) q_{i} \\
\mbox{s.t.} &&q\in \mathbb{Q}^{S}:=\left\{ q\in
\mathbb{R}_{+}^{s}\mid p_{i}(q_{i})\geq w(q_{S})\text{ for all
}i\in S\right\} .
\end{eqnarray*}%
The reader may notice that $\mathbb{Q}^{S}\ $is a nonempty set since $%
(q_{i}^{\{i\}})_{i\in S}\in \mathbb{Q}^{S}$, where $q_{i}^{\{i\}}$\ is an
optimal solution for retailer $i$'s RS-problem $(c,w,p_{i})$. Besides, $%
\mathbb{Q}^{S}\subset \prod_{i\in S}[0,q_{i}^{\ast }]$, where the order size
$q_{i}^{\ast }$\ satisfies $p_{i}(q_{i}^{\ast })=c$, and $\mathbb{Q}^{S}$ is
closed. Hence, there exists an optimal solution for this optimization
problem. Let $q_{i}^{S}$\ denote the optimal order size for retailer $i$\
when cooperating in coalition $S$.

A second type of cooperation is cooperation among a group of retailers $S$
and the supplier. The joint benefit of this coalition $S_{0}$ is%
\begin{equation*}
\sum_{i\in S}\Pi _{i}^{ret}\left( q_{i};w(q_{S})\right) +\sum_{i\in S}\Pi
_{i}^{sup}\left( q_{i};w(q_{S})\right) =\sum_{i\in S}\left(
p_{i}(q_{i})-c\right) q_{i},
\end{equation*}%
which shows a reduction in cost inefficiency for the companies due to the
absence of the intermediate wholesale prices. Under cooperation this
coalition optimizes%
\begin{eqnarray*}
\max  &&\sum_{i\in S}\left( p_{i}(q_{i})-c\right) q_{i} \\
\mbox{s.t.} &&q\in \mathbb{Q}_{c}^{S}:=\left\{ q\in
\mathbb{R}_{+}^{s}\mid p_{i}(q_{i})\geq c\text{ for all }i\in
S\right\} ,
\end{eqnarray*}%
which is equivalent to%
\begin{eqnarray}
\text{for all }i\in S:\qquad \max  &&\left( p_{i}(q_{i})-c\right) q_{i}
\label{IndividualOptimization} \\
\mbox{s.t.} &&p_{i}(q_{i})\geq c  \notag
\end{eqnarray}%
Similar arguments as above assure the existence of optimal
solutions for these optimization problems. Let $q_{i}^{c}$\ denote
the optimal order size for retailer $i$\ in this situation. Notice
that this quantity does not depend on $S$\ or $S_{0}$, since the
optimization problem (\ref{IndividualOptimization}) for $i$\ only
depends on individual parameters. Also, in general $q_{i}^{c}\neq
q_{i}^{S}$\ for all coalitions $S$\ with $i\in S$.

The profit functions arising from cooperation have nice properties, as
stated in the lemma below.

\begin{lemma}
\label{propi} Let $(N_{0},c,w,P)$ be an RS-situation, and $i\in S\subseteq N$
Then

\begin{enumerate}
\item[\emph{(P1)}] $\Pi _{i}^{ret}(q_{i}^{S};c)=\Pi _{i}^{ret}\left(
q_{i}^{S};w(q_{S}^{S})\right) +\Pi _{i}^{sup}(q_{i}^{S};w(q_{S}^{S}));$

\item[\emph{(P2)}] $\Pi _{i}^{ret}\left( q_{i}^{c};c\right) \geq \Pi
_{i}^{ret}\left( q_{i}^{S};c\right) ;$

\item[\emph{(P3)}] $\Pi _{i}^{ret}\left( q_{i}^{c};c\right) >\Pi
_{i}^{ret}\left( q_{i}^{S};w(q_{S}^{S})\right) $ and $\Pi _{i}^{ret}\left(
q_{i}^{c};c\right) >\Pi _{i}^{sup}\left( q_{i}^{S};w\left( q_{S}^{S}\right)
\right) .$
\end{enumerate}
\end{lemma}

\begin{proof}
(P1) follows immediately from the definitions of $\Pi _{i}^{ret}$ and $\Pi
_{i}^{sup}$. (P2) follows from $\Pi _{i}^{ret}\left( q_{i}^{c};c\right)
=\max_{q_{i}}\left( p_{i}(q_{i})-c\right) q_{i}\geq \left(
p_{i}(q_{i}^{S})-c\right) q_{i}^{S}$. Finally, (P3) follows from (P1), (P2),
$\Pi _{i}^{ret}\left( q_{i}^{S};w(q_{S}^{S})\right) >0$ and $\Pi
_{i}^{sup}\left( q_{i}^{S};w\left( q_{S}^{S}\right) \right) >0$.
\end{proof}

\bigskip Next we define the cooperative game corresponding to an
RS-situation. Its characteristic function is based on the maximum profit
that each coalition can reach.

\begin{definition}
Let $(N_{0},c,w,P)$ be an RS-situation. The corresponding RS-game $(N_{0},v)$
is defined by%
\begin{equation*}
v(S)=\sum_{i\in S}\Pi _{i}^{ret}\left( q_{i}^{S};w\left( q_{S}^{S}\right)
\right)
\end{equation*}%
and%
\begin{equation*}
v(S_{0})=\sum_{i\in S}\Pi _{i}^{ret}\left( q_{i}^{c};c\right)
\end{equation*}%
for all coalitions $S\subseteq N$, and $v(\emptyset )=0.$
\end{definition}

The definition of RS-situations and their corresponding games focuses on
retailer revenues arising from selling goods to consumers. As a consequence
one may interpret the game value $v(T)$ of a coalition $T$ as being \
`zero-normalised' with respect to the supplier. This explains why $%
v(\{0\})=0 $.

The definition above shows that a coalition of retailers benefits from a
lower wholesale price per unit while a coalition including the supplier
increases its profit due to the absence of the intermediate wholesale
prices. This provides the companies in the chain with sufficient incentives
for cooperation. Also, cooperation with the supplier is attractive for
retailers since $v(S_{0})>v(S)$ by property (P3). The reader may notice that
the supplier does not contribute with a fixed quantity of gain but rather he
reduces the cost inefficiency by suppressing the intermediate wholesale
prices. To be more precise, if the supplier joins a coalition $S$ of
retailers then this leads to an increase of
\begin{equation*}
v(S_{0})-v(S)=\sum_{i\in S}\left( p_{i}(q_{i}^{c})-c\right)
q_{i}^{c}-\sum_{i\in S}\left( p_{i}(q_{i}^{S})-w(q_{S}^{S})\right) q_{i}^{S}.
\end{equation*}%
Besides the optimal quantity for each retailer \emph{does }depend upon the
presence of the supplier in the coalition. Namely, if the supplier is
present then the optimal quantity for referee $i$ is $q_{i}^{c}$ and if the
supplier is not present then it is $q_{i}^{S}$. See for instance example 5.7
in which $q_{i}^{S}=49$ (the supplier is not present) and $q_{i}^{c}=48\frac{%
1}{5}$ (the supplier is present)\ for $i=1,2$ and $S\subseteq N.$ Hence, the
difference among $v(S_{0})$ and $v(S)$ does not just depend upon the gain of
the supplier. The increase is also due to the retailers: cooperation
increases the gain of the retailers since for all $i\in S$%
\begin{equation*}
\left( p_{i}(q_{i}^{c})-c\right) q_{i}^{c}>\left(
p_{i}(q_{i}^{S})-w(q_{S}^{S})\right) q_{i}^{S}.
\end{equation*}%
Therefore, the supplier has reasons to share the gain from cooperation with
the retailers.

The example below shows an RS-situation and its corresponding RS-game.

\begin{example}
\label{Ex_1}Let $(N_{0},c,w,P)$ be an RS-situation with $N_{0}=\{0,1,2\}$, $%
c=2$, $p_{1}(q)=7-q$, $p_{2}(q)=8-q$ and
\begin{equation*}
w(q)=\left\{
\begin{array}{ll}
5, & \frac{1}{4}\leq q\leq 1, \\
2+3/q, & q>1.%
\end{array}%
\right.
\end{equation*}%
The optimal order sizes are $q_{1}^{S}=q_{1}^{c}=2\frac{1}{2}$ and $%
q_{2}^{S}=q_{2}^{c}=3$ for all $S\subseteq N$. This implies the RS-game as
given in the table below.%
\begin{equation*}
\begin{tabular}{c||ccccccc}
\hline
$T$ & $\{0\}$ & $\{1\}$ & $\{2\}$ & $\{0,1\}$ & $\{0,2\}$ & $\{1,2\}$ & $%
\{0,1,2\}$ \\ \hline
$v(T)$ & $0$ & $3\frac{1}{4}$ & $6$ & $6\frac{1}{4}$ & $9$ & $12\frac{1}{4}$
& $15\frac{1}{4}$ \\ \hline
\end{tabular}%
\end{equation*}%
This game has positive values for coalitions $T\neq \{0\}$, is superadditive
and monotone increasing. In addition, $v(N_{0})=v(\{0,1\})+v(\{0,2\})$.
\end{example}

The properties that are observed in this example hold in general, as the
next lemma shows.

\begin{lemma}
\label{PropMonSupadd}Let $(N_{0},v)$ be an RS-game. Then

\begin{itemize}
\item[(i)] $v(T)>0$ for all coalitions $T\neq \{0\}$;

\item[(ii)] $v$ is superadditive;

\item[(iii)] $v$ is strict monotone increasing;

\item[(iv)] $v(S_{0})=\sum_{i\in S}v(\{0,i\})$ and $v(S_{0})-v(S_{0}%
\backslash \{i\})=v(\{0,i\})$ for all $S\subseteq N$ and $i\in S$.
\end{itemize}
\end{lemma}

\begin{proof}
$(i)$ The definition of the game $(N_{0},v)$ and the positive profit margins
for the retailers ($p_{i}(q)>w(q)$) imply that $v(T)>0$ for any coalition $%
T\neq \{0\}$.

$(ii)$ Let $S,T\subseteq N$ be two disjoint coalitions of retailers. By
definition of the RS-game
\begin{equation*}
v(S)+v(T)=\sum_{i\in S}\Pi _{i}^{ret}\left( q_{i}^{S};w\left(
q_{S}^{S}\right) \right) +\sum_{i\in T}\Pi _{i}^{ret}(q_{i}^{T};w\left(
q_{T}^{T}\right) ).
\end{equation*}%
Define the specific order quantity $\hat{q}_{i}$ for retailer $i$ in
coalition $S\cup T$ by $\hat{q}_{i}=q_{i}^{S}$ if $i\in S$ and $\hat{q}%
_{i}=q_{i}^{T}$ if $i\in T$. Now
\begin{eqnarray*}
\lefteqn{\sum_{i\in S}\Pi _{i}^{ret}\left( q_{i}^{S};w\left(
q_{S}^{S}\right) \right) +\sum_{i\in T}\Pi _{i}^{ret}(q_{i}^{T};w\left(
q_{T}^{T}\right) )} \\
&\leq &\sum_{i\in S}\Pi _{i}^{ret}\left( \hat{q}_{i};w\left( \hat{q}_{S\cup
T}\right) \right) +\sum_{i\in T}\Pi _{i}^{ret}\left( \hat{q}_{i};w\left(
\hat{q}_{S\cup T}\right) \right)
\end{eqnarray*}%
because this larger coalition enjoys a lower wholesale price than before: $w(%
\hat{q}_{S\cup T})=w(q_{S}^{S}+q_{T}^{T})\leq \min \{w\left(
q_{S}^{S}\right) ,w\left( q_{T}^{T}\right) \}$. Finally,%
\begin{eqnarray*}
\lefteqn{\sum_{i\in S}\Pi _{i}^{ret}\left( \hat{q}_{i};w\left( \hat{q}%
_{S\cup T}\right) \right) +\sum_{i\in T}\Pi _{i}^{ret}\left( \hat{q}%
_{i};w\left( \hat{q}_{S\cup T}\right) \right) } \\
&=&\sum_{i\in S\cup T}\Pi _{i}^{ret}\left( \hat{q}_{i};w\left( \hat{q}%
_{S\cup T}\right) \right) \\
&\leq &\sum_{i\in S\cup T}\Pi _{i}^{ret}\left( q_{i}^{S\cup T};w\left(
q_{S\cup T}^{S\cup T}\right) \right) =v(S\cup T),
\end{eqnarray*}%
since the quantities $\hat{q}_{i}$ need not be optimal for coalition $S\cup
T $.

Furthermore,%
\begin{eqnarray*}
v(S_{0})+v(T) &=&\sum_{i\in S}\Pi _{i}^{ret}\left( q_{i}^{c};c\right)
+\sum_{i\in T}\Pi _{i}^{ret}(q_{i}^{T};w\left( q_{T}^{T}\right) ) \\
&<&\sum_{i\in S}\Pi _{i}^{ret}\left( q_{i}^{c};c\right) +\sum_{i\in T}\Pi
_{i}^{ret}(q_{i}^{c};c) \\
&=&\sum_{i\in S\cup T}\Pi _{i}^{ret}\left( q_{i}^{c};c\right) =v(S_{0}\cup
T),
\end{eqnarray*}%
where the second inequality is due to property (P3).

$(iii)$ This follows immediately from $(i)$, $(ii)$ and (P3).

$(iv)$ These results follow directly from the definition of the game.
\end{proof}

\bigskip The fourth property in this lemma shows that each optimal profit of
a large coalition including the supplier is composed of optimal profits for
pairs of the supplier and a retailer. This is due to the fact that the
retailers do not compete for customers but rather serve their own market.

\section{The core of RS-games\label{SecBalancedness}}

The core is an important set-solution for cooperative games. In this section
we show that every RS-game is balanced, that is, its core is nonempty. But
first we present the core structure for RS-games.

\begin{theorem}
\label{ThCoreReduced}%\mbox{}\newline
Let $(N_{0},c,w,P)$ be an RS-situation and $(N_{0},v)$ the corresponding
RS-game. The core of this game equals
\begin{equation*}
Core(N_{0},v)=\left\{ x\in \mathbb{R}^{n_{0}}\left\vert
\begin{array}{l}
\sum_{i\in N_{0}}x_{i}=v(N_{0});~x_{i}\leq v(\{0,i\}),~i\in N; \\
\sum_{i\in S}x_{i}\geq v(S),~S\subseteq N%
\end{array}%
\right. \right\} .
\end{equation*}
\end{theorem}

\begin{proof}
Let $i\in N$ and $x\in Core(N_{0},v)$. The core conditions $\sum_{i\in
N_{0}}x_{i}=v(N_{0})$ and $\sum_{j\in N_{0}\setminus \{i\}}x_{j}\geq
v(N_{0}\setminus \{i\})$ imply $x_{i}\leq v(N_{0})-v(N_{0}\setminus \{i\})$,
which leads to%
\begin{equation*}
x_{i}\leq v(N_{0})-v(N_{0}\setminus \{i\})=v(N_{0})-\left(
v(N_{0})-v(\{0,i\})\right) =v(\{0,i\}),
\end{equation*}%
in which the first equality follows from property (iv) in lemma \ref%
{PropMonSupadd}.

We proceed by showing that the core conditions $x_{0}+\sum_{i\in S}x_{i}\geq
v(S_{0})$, $S\subseteq N$, are superfluous. We obtain subsequently%
\begin{eqnarray*}
x_{0}+\sum_{i\in S}x_{i} &=&\sum_{i\in N_{0}}x_{i}-\sum_{j\in N\setminus
S}x_{j} \\
&\geq &v(N_{0})-\sum_{j\in N\setminus S}v(\{0,j\}) \\
&=&\sum_{j\in S}v(\{0,j\})=v(S_{0}),
\end{eqnarray*}%
by using respectively $\sum_{i\in N_{0}}x_{i}=v(N_{0})$, $x_{j}\leq
v(\{0,j\})$ and property (iv) in lemma \ref{PropMonSupadd}.
\end{proof}

\bigskip This theorem says that the stability conditions for coalitions
including the supplier (i.e. $\sum_{i\in S_{0}}x_{i}\geq v(S_{0})$) can be
replaced by the conditions $x_{i}\leq v(\{0,i\})$ for all retailers $i.$
This allows for an easier expression of the core of RS-games.

Using the above theorem, the core of example \ref{Ex_1} can be expressed as%
\begin{equation*}
Core(N_{0},v)=\left\{ (x_{0},x_{1},x_{2})\left\vert
\begin{array}{l}
3\frac{1}{4}\leq x_{1}\leq 6\frac{1}{4},~6\leq x_{2}\leq 9, \\
x_{1}+x_{2}\geq 12\frac{1}{4},~x_{0}+x_{1}+x_{2}=15\frac{1}{4}%
\end{array}%
\right. \right\} .
\end{equation*}%
One immediately sees that this core is nonempty since $(0,6\frac{1}{4},9)\in
Core(N_{0},v)$. Nonemptiness of the core of RS-games in general is shown in
the theorem below.

\begin{theorem}
\label{ThBalanced}Let $(N_{0},c,w,P)$ be an RS-situation and $(N_{0},v)$ the
corresponding RS-game. Then $(N_{0},v)$ is balanced.
\end{theorem}

\begin{proof}
Define the allocation $x^{a}(v)$ by $x_{0}^{a}(v)=0$ and $x_{i}^{a}(v)=\Pi
_{i}^{ret}\left( q_{i}^{c};c\right) =v(\{0,i\})$, $i\in N$. In this
allocation the retailer receives all the benefit from cooperation with the
supplier, while the supplier receives nothing. First notice that%
\begin{equation*}
\sum_{i\in N_{0}}x_{i}^{a}(v)=\sum_{i\in N}\Pi _{i}^{ret}\left(
q_{i}^{c};c\right) =v(N_{0}).
\end{equation*}%
Hence, $x^{a}(v)$ is an efficient allocation. Next, consider a coalition $%
S\subseteq N$. Then%
\begin{equation*}
\sum_{i\in S}x_{i}^{a}(v)=\sum_{i\in S}\Pi _{i}^{ret}\left(
q_{i}^{c};c\right) >\sum_{i\in S}\Pi _{i}^{ret}\left(
q_{i}^{S};w(q_{S}^{S})\right) =v(S),
\end{equation*}%
where the inequality follows from property (P3). Finally, by definition $%
x_{i}^{a}(v)\leq v(\{0,i\})$. We conclude that $x^{a}(v)\in Core(N_{0},v)$,
thus, the game is balanced.
\end{proof}

\bigskip The proof of this theorem shows that the allocation $%
x^{a}(v)=\left( 0,\left( v(\{0,i\}\right) _{i\in N}\right) $\ always belongs
to the core of an RS-game. This allocation will be called the altruistic
allocation since it is the worst possible core-allocation for the supplier,
namely the only one in which he receives nothing.

As a corollary of this theorem we obtain balancedness of subgames
including the supplier in the player set.

%, that is, subgames in which the reduced player set includes the
%supplier.

\begin{corollary}
Let $(N_{0},c,w,P)$ be an RS-situation and $S$ a coalition of retailers. Let
$P_{S}$ denote the restriction of the vector of consumer price functions $P$
to coalition $S$. Then the game $(S_{0},v_{S_{0}})$ corresponding to the
RS-situation $(S_{0},c,w,P_{S})$ is balanced.
\end{corollary}

This corollary and its preceding theorem show that cooperation is profitable
for all companies in the distribution chain because (a) it results in higher
profits, and (b) there exists a core-allocation of the joint optimal profit,
that is, an allocation upon which no coalition can improve.

The core of RS-games has a nice alternative interpretation in terms of the
underlying distribution chain, as we will see shortly. For this, we need the
following lemma and its corollary.

\begin{lemma}
Let $(N_{0},c,w,P)$ be an RS-situation. Then there exist prices $w_{i}^{\ast
}\in \lbrack c,p_{i}(q_{i}^{c})]$, $i\in N$, such that%
\begin{equation*}
\Pi _{i}^{ret}\left( q_{i}^{c};w_{i}^{\ast }\right) \geq \max_{S\subseteq
N:i\in S}\Pi _{i}^{ret}\left( q_{i}^{S};w(q_{S}^{S})\right) .
\end{equation*}
\end{lemma}

\begin{proof}
Consider a coalition $S$ of retailers and let $i\in S$ be one of them. Then
by property (P3)
\begin{equation*}
\Pi _{i}^{ret}\left( q_{i}^{c};c\right) >\Pi _{i}^{ret}\left(
q_{i}^{S};w\left( q_{S}^{S}\right) \right) >0.
\end{equation*}%
This implies that there exists a number $\beta ^{S}$, $0<\beta ^{S}<\Pi
_{i}^{ret}(q_{i}^{c};c)$, such that $\Pi _{i}^{ret}\left( q_{i}^{S};w\left(
q_{S}^{S}\right) \right) =\Pi _{i}^{ret}\left( q_{i}^{c};c\right) -\beta
^{S} $. Furthermore, there is a wholesale price $w_{i}^{S}\in \lbrack
c,p_{i}(q_{i}^{c})]$ such that $\Pi
_{i}^{sup}(q_{i}^{c};w_{i}^{S})=(w_{i}^{S}-c)q_{i}^{c}\leq \beta ^{S}$. Thus%
\begin{eqnarray*}
\Pi _{i}^{ret}\left( q_{i}^{S};w\left( q_{S}^{S}\right) \right) &=&\Pi
_{i}^{ret}\left( q_{i}^{c};c\right) -\beta ^{S} \\
&\leq &\Pi _{i}^{ret}\left( q_{i}^{c};c\right) -\Pi
_{i}^{sup}(q_{i}^{c};w_{i}^{S}) \\
&=&\Pi _{i}^{ret}\left( q_{i}^{c};w_{i}^{S}\right) ,
\end{eqnarray*}%
where property (P1) is used in the last equality. Let $w_{i}^{\ast
}=\min_{S\ni i}w_{i}^{S}$ be the lowest wholesale price for retailer $i$.
Then%
\begin{equation*}
\Pi _{i}^{ret}\left( q_{i}^{c};w_{i}^{\ast }\right) \geq \Pi
_{i}^{ret}\left( q_{i}^{c};w_{i}^{S}\right) \geq \Pi _{i}^{ret}\left(
q_{i}^{S};w(q_{S}^{S})\right)
\end{equation*}%
for all $S\subseteq N$ with $i\in S$, which concludes the proof.
\end{proof}

\bigskip Notice that this lemma generates upper bounds lower than $%
p_{i}(q_{i}^{c})$ for the values $w_{i}^{\ast }$. Lower bounds for $%
w_{i}^{\ast }$\ are given by $c$. As a corollary we obtain the following
weaker result.

\begin{corollary}
\label{ws}Let $(N_{0},c,,w,P)$ be an RS-situation. Then there exist prices $%
w_{i}^{\ast }\geq c$, $i\in N$, such that
\begin{equation*}
\sum_{i\in S}\Pi _{i}^{ret}\left( q_{i}^{c};w_{i}^{\ast }\right) \geq
\sum_{i\in S}\Pi _{i}^{ret}\left( q_{i}^{S};w(q_{S}^{S})\right) =v(S),
\end{equation*}%
or equivalently,%
\begin{equation*}
\sum_{i\in S}q_{i}^{c}w_{i}^{\ast }\leq \sum_{i\in
S}p_{i}(q_{i}^{c})q_{i}^{c}-v(S),
\end{equation*}%
for all coalitions $S$ of retailers.
\end{corollary}

Using this corollary we provide an alternative formulation of the core of an
RS-game.

\begin{theorem}
\label{CoreStructure}Let $(N_{0},c,w,P)$ be an RS-situation and $(N_{0},v)$
the corresponding RS-game. Then $x\in Core(N_{0},v)$ if and only if
\begin{equation*}
x_{0}=\sum_{i\in N}\Pi _{i}^{sup}\left( q_{i}^{c};w_{i}^{\ast }\right) \text{
and }x_{i}=\Pi _{i}^{ret}\left( q_{i}^{c};w_{i}^{\ast }\right) \text{, }i\in
N,
\end{equation*}%
for some $w^{\ast }\in \mathbb{R}^{n}$ that satisfies corollary \ref{ws}.
\end{theorem}

\begin{proof}
First, according to theorem \ref{ThBalanced}$\ Core(N_{0},v)\neq \emptyset $
All elements $x\in Core(N_{0},v)$ satisfy%
\begin{equation*}
v(\{i\})\leq x_{i}\leq v(N_{0})-v(N_{0}\setminus \{i\})
\end{equation*}%
for all $i\in N_{0}$. Using the definition of the RS-game this condition is
equivalent to%
\begin{equation*}
x_{i}^{L}=\Pi _{i}^{ret}\left( q_{i}^{i};w(q_{i}^{\{i\}})\right) \leq
x_{i}\leq \Pi _{i}^{ret}\left( q_{i}^{c};c\right) =x_{i}^{H}
\end{equation*}%
for all retailers $i\in N$. Now, let $x^{\ast }=(x_{0}^{\ast },x_{1}^{\ast
},...,x_{n}^{\ast })\in Core(N_{0},v)$. Notice that $\Pi _{i}^{ret}\left(
q_{i}^{c};p_{i}(q_{i}^{c})\right) =0$ and so%
\begin{equation*}
x_{i}^{\ast }\in \lbrack x_{i}^{\ast L},x_{i}^{\ast H}]\subset \lbrack \Pi
_{i}^{ret}\left( q_{i}^{c};p_{i}(q_{i}^{c})\right) ,\Pi _{i}^{ret}\left(
q_{i}^{c};c\right) ]
\end{equation*}%
for all $i\in N$. Because of this, there exists $w_{i}^{\ast }\in \lbrack
c,p_{i}(q_{i}^{c})]$ such that $x_{i}^{\ast }=\Pi _{i}^{ret}\left(
q_{i}^{c};w_{i}^{\ast }\right) $. The efficiency of core-elements implies $%
x_{0}^{\ast }=\sum_{i\in N}\Pi _{i}^{sup}\left( q_{i}^{c};w_{i}^{\ast
}\right) $. For any coalition $S\subseteq N$ of retailers%
\begin{equation*}
v(S)\leq \sum_{i\in S}x_{i}^{\ast }\Longleftrightarrow \sum_{i\in S}\Pi
_{i}^{ret}\left( q_{i}^{S};w(q_{S}^{S})\right) \leq \sum_{i\in S}\Pi
_{i}^{ret}\left( q_{i}^{c};w_{i}^{\ast }\right) .
\end{equation*}%
Hence, the $w_{i}^{\ast }$ satisfy corollary \ref{ws}. This concludes the
first part of the proof.

Second, let $w^{\ast }\in \mathbb{R}^{n}$ satisfy corollary \ref{ws}. Define
the allocation $x^{\ast }$ by%
\begin{equation*}
x_{i}^{\ast }=\left\{
\begin{array}{ll}
\sum_{j\in N}\Pi _{j}^{sup}\left( q_{j}^{c};w_{j}^{\ast }\right) , & i=0, \\
\Pi _{i}^{ret}\left( q_{i}^{c};w_{i}^{\ast }\right) , & i\in N.%
\end{array}%
\right.
\end{equation*}%
Notice first that%
\begin{equation*}
\sum_{i\in N_{0}}x_{i}^{\ast }=\sum_{i\in N}\Pi _{i}^{ret}\left(
q_{i}^{c};w_{i}^{\ast }\right) +\sum_{j\in N}\Pi _{j}^{sup}\left(
q_{j}^{c};w_{j}^{\ast }\right) =v(N_{0})
\end{equation*}%
where the last equality is due to property (P1). Hence, $x^{\ast }$ is an
efficient allocation. Next, consider a coalition $S\subseteq N$. Then%
\begin{equation*}
\sum_{i\in S}x_{i}^{\ast }=\sum_{i\in S}\Pi _{i}^{ret}\left(
q_{i}^{c};w_{i}^{\ast }\right) \geq \sum_{i\in S}\Pi _{i}^{ret}\left(
q_{i}^{S};w(q_{S}^{S})\right) =v(S),
\end{equation*}%
where the inequality follows from corollary \ref{ws}. Further,%
\begin{eqnarray*}
\sum_{i\in S_{0}}x_{i}^{\ast } &=&\sum_{i\in S}\Pi _{i}^{ret}\left(
q_{i}^{c};w_{i}^{\ast }\right) +\sum_{i\in N}\Pi _{i}^{sup}\left(
q_{i}^{c};w_{i}^{\ast }\right) \\
&=&\sum_{i\in S}\Pi _{i}^{ret}\left( q_{i}^{c};c\right) +\sum_{i\in
N\setminus S}\Pi _{i}^{sup}\left( q_{i}^{c};w_{i}^{\ast }\right) \\
&\geq &\sum_{i\in S}\Pi _{i}^{ret}\left( q_{i}^{c};c\right) =v(S_{0}),
\end{eqnarray*}%
where the second equality is due to property (P1). We conclude that $x^{\ast
}\in Core(N_{0},v)$.
\end{proof}

\bigskip This theorem shows that for each core-allocation $x$\ there exist
fixed wholesale prices $w^{\ast }$\ such that the retailer's share $x_{i}$\
corresponds to its optimal profit under cooperation with the supplier given
the wholesale price $w_{i}^{\ast }$, $x_{i}=\Pi _{i}^{ret}\left(
q_{i}^{c};w_{i}^{\ast }\right) $, and the supplier's share equals $%
x_{0}=\sum_{i\in N}\Pi _{i}^{sup}\left( q_{i}^{c};w_{i}^{\ast }\right) $.
Hence, each core-allocation has a natural interpretation in terms of the
underlying distribution chain.

Further, this theorem shows that the structure of the core is determined by
the values $w_{i}^{\ast }.$ Corollary \ref{ws} generates for each coalition $%
S$ of retailers an upper bound for $\{w_{i}^{\ast }\}_{i\in S}$, which is
related to the core-condition of this coalition. The lower bound $%
w_{i}^{\ast }=c$ corresponds to the upper bound $v(\{0,i\})$ for $x_{i}$.
This corresponds with the formulation of the core in theorem \ref%
{ThCoreReduced}. The example below illustrates these observations.

\begin{example}
\label{Ex_3}Let $(N_{0},c,w,P)$ be the RS-situation with $N_{0}=\{0,1,2\}$, $%
c=1\frac{4}{5}$, $p_{1}(q)=p_{2}(q)=50-\frac{q}{2}$, and

\begin{equation*}
w(q)=\left\{
\begin{array}{ll}
11, & 0\leq q\leq 10, \\
1+\frac{100}{q}, & 10<q\leq 100, \\
2, & q>100.%
\end{array}%
\right.
\end{equation*}%
The optimal order sizes are $q_{i}^{S}=49$ and $q_{i}^{c}=48\frac{1}{5}$ for
$i=1,2$ and $S\subseteq N.$ The coalitional values in the corresponding game
are
\begin{equation*}
\begin{tabular}{c||ccccccc}
\hline
$T$ & $\{0\}$ & $\{1\}$ & $\{2\}$ & $\{0,1\}$ & $\{0,2\}$ & $\{1,2\}$ & $%
\{0,1,2\}$ \\ \hline
$v(T)$ & $0$ & $1100\frac{1}{2}$ & $1100\frac{1}{2}$ & $1161\frac{31}{50}$ &
$1161\frac{31}{50}$ & $2301$ & $2323\frac{6}{25}$ \\ \hline
\end{tabular}%
\end{equation*}%
The core of this game equals%
\begin{eqnarray*}
Core(N_{0},v) &=&\left\{ (x_{0},x_{1},x_{2})\left\vert
\begin{array}{l}
1100\frac{1}{2}\leq x_{i}\leq 1161\frac{31}{50},~i=1,2; \\
x_{1}+x_{2}\geq 2301;~x_{0}+x_{1}+x_{2}=2323\frac{6}{25}%
\end{array}%
\right. \right\} \\
&=&\left\{ \left. \left(
\begin{array}{c}
48\frac{1}{5}\left( w_{1}^{\ast }+w_{2}^{\ast }\right) -173\frac{13}{25}, \\
1248\frac{19}{50}-48\frac{1}{5}w_{1}^{\ast }, \\
1248\frac{19}{50}-48\frac{1}{5}w_{2}^{\ast }%
\end{array}%
\right) \right\vert
\begin{array}{l}
1\frac{4}{5}\leq w_{i}^{\ast }\leq 3\frac{82}{1205},~i=1,2; \\
w_{1}^{\ast }+w_{2}^{\ast }\leq 4\frac{74}{1205}.%
\end{array}%
\right\} .
\end{eqnarray*}%
The correspondence between the two formulations is clear:%
\begin{equation*}
1\frac{4}{5}\leq w_{i}^{\ast }\Longleftrightarrow x_{i}\leq 1248\frac{19}{50}%
-48\frac{1}{5}\cdot 1\frac{4}{5}=1161\frac{31}{50}
\end{equation*}%
and%
\begin{equation*}
w_{i}^{\ast }\leq 3\frac{82}{1205}\Longleftrightarrow x_{i}\geq 1248\frac{19%
}{50}-48\frac{1}{5}\cdot 3\frac{82}{1205}=1100\frac{1}{2}
\end{equation*}%
for $i=1,2$. Also,
\begin{equation*}
w_{1}^{\ast }+w_{2}^{\ast }\leq 4\frac{74}{1205}\Longleftrightarrow
x_{1}+x_{2}\geq 2496\frac{19}{25}-48\frac{1}{5}\cdot 4\frac{74}{1205}=2301.
\end{equation*}%
Finally, the equality%
\begin{equation*}
48\frac{1}{5}\left( w_{1}^{\ast }+w_{2}^{\ast }\right) -173\frac{13}{25}+1248%
\frac{19}{50}-48\frac{1}{5}w_{1}^{\ast }+1248\frac{19}{50}-48\frac{1}{5}%
w_{2}^{\ast }=2323\frac{6}{25}
\end{equation*}%
implies that the allocations are efficient.
\end{example}

This example shows that there is a one-to-one relation between the
conditions of the core of RS-games in both the reduced formulation in
theorem \ref{ThCoreReduced}\ and the alternative formulation in theorem \ref%
{CoreStructure}.

\section{A solution for RS-games\label{SecSolution}}

In the previous section one single-valued solution for RS-games was already
discussed briefly, namely the altruistic allocation $x^{a}(v)$. This
allocation always belongs to the core but it is not fair because it assigns
a zero payoff to the supplier although this company is needed to obtain the
largest total profits. Instead, a suitable solution for RS-games should
assign a positive payoff to the supplier and it should belong to the core of
the game. Four desirable properties for a single-valued solution $\varphi $
for RS-games $(N_{0},v)$\ are:

\begin{description}
\item[(EF)] Efficiency. $\sum_{i\in N_{0}}\varphi _{i}(v)=v(N_{0}).$

\item[(SR)] Stability for retailers. $\sum_{i\in S}\varphi _{i}(v)\geq v(S)$
for all coalitions $S\subseteq N.$

\item[(RR)] Retailer reduction. $\varphi _{i}(v)=v(\{0,i\})-\frac{%
v(S_{0}^{i})-v(S^{i})}{s^{i}}$ for some coalition $S^{i}\subseteq N$, for
all $i\in N.$

\item[(PD)] Preservation of differences for retailers. $\varphi
_{i}(v)-\varphi _{j}(v)=v(\{0,i\})-v(\{0,j\})$ for all $i,j\in N$ with $%
i\neq j.$
\end{description}

\noindent \textit{Efficiency} implies that the total benefit is divided
among the players, while \textit{stability for retailers }ensures
coalitional stability for all coalitions of retailers. The \textit{retailer
reduction }property says that a retailer receives an amount smaller than his
joint profit with the supplier $v(\{0,i\})$. The reduction equals the gain
per capita for coalition $S^{i}$ from cooperation with the supplier, $%
(v(S_{0}^{i})-v(S^{i}))/s^{i}$, for some coalition of retailers $S^{i}$.
Finally, the \textit{preservation of differences for retailers} property is
a modification of the \textit{preservation of differences} property by \cite%
{HM89}. The (PD) property states that the difference in payoffs for two
retailers should equal the difference in their joint profits with the
supplier.

The main result in this section states that there exists a unique solution
for RS-games satisfying the properties (EF), (SR), (RR) and (PD).

\begin{theorem}
Let $(N_{0},c,w,P)$ be an RS-situation and $(N_{0},v)$ the corresponding
RS-game. The unique solution $\xi $ on the class of RS-games, $RS^{N_{0}}$,
satisfying (EF), (SR), (RR) and (PD) is $\xi (v)=\left( \xi _{i}(v)\right)
_{i\in N_{0}}$ defined by
\begin{equation*}
\xi _{i}(v)=\left\{
\begin{array}{ll}
n\beta , & i=0, \\
v(\{0,i\})-\beta , & i\in N,%
\end{array}%
\right.
\end{equation*}%
where $\beta =\min_{S\subseteq N,S\neq \emptyset }\left\{ \frac{v(S_{0})-v(S)%
}{s}\right\} .$
\end{theorem}

\begin{proof}
It is clear that $\xi (v)$ satisfies (EF), (SR), (RR) and (PD).

To show the converse, take a solution $\varphi $ on the class of RS-games
that satisfies (EF), (SR), (RR) and (PD). By (RR), $\varphi
_{i}(v)=v(\{0,i\})-\alpha _{i}$ with $\alpha
_{i}=(v(S_{0}^{i})-v(S^{i}))/s^{i}$ for some coalition $S^{i}\subseteq N$,
for all retailers $i$. By (PD), $\alpha _{i}=\alpha _{j}$ for all $i,j\in N$
with $i\neq j$. This implies $\alpha _{i}=\alpha _{\ast }$ for all $i\in N$.
According to (SR) $\sum_{i\in S}\varphi _{i}(v)=v(S_{0})-s\alpha _{\ast
}\geq v(S)$ or equivalently $\alpha _{\ast }\leq (v(S_{0})-v(S))/s$ for all
coalitions $S\subseteq N$. But then $\alpha _{\ast }=\min_{S\subseteq
N,S\neq \emptyset }\{(v(S_{0})-v(S))/s\}$. Finally, by (EF) we conclude\ $%
\varphi =\xi $.
\end{proof}

\bigskip This unique solution $\xi $ is called the minimal-gain-per-capita
solution (in short:\ mgpc-solution) because each retailer pays the minimal
gain per capita $\beta $ to the supplier. Two properties of mgpc-solutions
follow.

\begin{lemma}
For all RS-games $(N_{0},v)$ the mgpc-solution is a core allocation, $\xi
(v)\in Core(N_{0},v)$, and it assures a positive payoff to the supplier, $%
\xi _{0}(v)>0$.
\end{lemma}

\begin{proof}
By property \textit{(iii)} in lemma \ref{PropMonSupadd} $v(S_{0})>v(S)$,
which implies $\beta >0$. Hence,%
\begin{equation*}
\xi _{i}(v)=v(\{0,i\})-\beta <v(\{0,i\})
\end{equation*}%
for all retailers $i$. Together with (EF) and (SR) we conclude $\xi (v)\in
Core(N_{0},v)$. Second, the positive value of $\beta $ immediately implies $%
\xi _{0}(v)>0$.
\end{proof}

\bigskip This lemma shows that the mgpc-solution $\xi (v)$ is a stable
allocation since it belongs to the core. Further, the supplier prefers this
allocation to the altruistic allocation $x^{a}(v)$\ because it assigns a
larger payoff to him.

Upon comparison of the mgpc-solution with the Shapley value $Sh(v)$ \cite%
{SH53}, we observe the following. In the table below the solutions are
mentioned for example \ref{Ex_1}.%
\begin{equation*}
\begin{tabular}{c|c|c}
$\xi (v)$ & $x^{a}(v)$ & $Sh(v)$ \\ \hline
&  &  \\
$\left(
\begin{array}{c}
3 \\
4\frac{3}{4} \\
7\frac{1}{2}%
\end{array}%
\right) $ & $\left(
\begin{array}{c}
0 \\
6\frac{1}{4} \\
9%
\end{array}%
\right) $ & $\left(
\begin{array}{c}
2 \\
5\frac{1}{4} \\
8%
\end{array}%
\right) $%
\end{tabular}%
\end{equation*}%
In this example the Shapley value belongs to the core since the game is
convex. The supplier prefers the mgpc-solution to the Shapley value since it
results in a larger payoff 3 instead of 2. If in this example the wholesale
price function $w$ is changed to $w(q)=5$ if $0\leq q\leq 1$ and $%
w(q)=9/2+1/(2q)$ if $q>1$ then the solutions change as follows.%
\begin{equation*}
\begin{tabular}{c|c|c}
$\xi (v)$ & $x^{a}(v)$ & $Sh(v)$ \\ \hline
&  &  \\
$\left(
\begin{array}{c}
10\frac{3}{8} \\
1\frac{1}{16} \\
3\frac{13}{16}%
\end{array}%
\right) $ & $\left(
\begin{array}{c}
0 \\
6\frac{1}{4} \\
9%
\end{array}%
\right) $ & $\left(
\begin{array}{c}
5\frac{31}{48} \\
3\frac{71}{96} \\
5\frac{83}{96}%
\end{array}%
\right) $%
\end{tabular}%
\end{equation*}%
Again, all three solutions belong to the core and the supplier prefers the
mgpc-solution. In example \ref{Ex_3} the solutions are as follows.%
\begin{equation*}
\begin{tabular}{c|c|c}
$\xi (v)$ & $x^{a}(v)$ & $Sh(v)$ \\ \hline
&  &  \\
$\left(
\begin{array}{c}
22\frac{6}{25} \\
1150\frac{1}{2} \\
1150\frac{1}{2}%
\end{array}%
\right) $ & $\left(
\begin{array}{c}
0 \\
1161\frac{31}{50} \\
1161\frac{31}{50}%
\end{array}%
\right) $ & $\left(
\begin{array}{c}
27\frac{59}{75} \\
1147\frac{109}{150} \\
1147\frac{109}{150}%
\end{array}%
\right) $%
\end{tabular}%
\end{equation*}%
Here, the Shapley value is not a core-allocation. Obviously, the supplier
prefers the mgpc-solution to the altruistic allocation. From these
observations we conclude that the mgpc-solution $\xi $ is suitable for
RS-games because (a) it recognizes the importance of the supplier in
achieving full cooperation and (b) it always belongs to the core of the
RS-game, as opposed to the Shapley value.

Finally, to conclude this section, the four examples below show that the
properties (EF), (SR), (RR) and (PD) are logically independent.

\begin{example}
Consider the solution $\varphi $ on $RS^{N_{0}}$ defined by%
\begin{equation*}
\varphi _{i}(v)=\left\{
\begin{array}{ll}
0, & i=0, \\
v(\{0,i\})-\beta , & i\in N.%
\end{array}%
\right.
\end{equation*}%
{$\varphi $ satisfies (SR), (RR) and (PD) but not (EF).}
\end{example}

\begin{example}
Consider $\varphi $ on $RS^{N_{0}}$ defined by%
\begin{equation*}
\varphi _{i}(v)=\left\{
\begin{array}{ll}
n\beta ^{\ast }, & i=0, \\
v(\{0,i\})-\beta ^{\ast }, & i\in N,%
\end{array}%
\right.
\end{equation*}%
where $\beta ^{\ast }=\max_{S\subseteq N,S\neq \emptyset }\left\{
(v(S_{0})-v(S))/s\right\} $. {$\varphi $ satisfies (EF), (RR) and (PD) but
not (SR).}
\end{example}

\begin{example}
Consider $\varphi $ on $RS^{N_{0}}$ defined by%
\begin{equation*}
\varphi _{i}(v)=\left\{
\begin{array}{ll}
n\left( \beta -1\right) , & i=0, \\
v(\{0,i\})-(\beta -1), & i\in N.%
\end{array}%
\right.
\end{equation*}%
{$\varphi $ satisfies (EF), (SR) and (PD) but not (RR).}
\end{example}

\begin{example}
Consider $\varphi $ on $RS^{N_{0}}$ defined by%
\begin{equation*}
\varphi _{i}(v)=\left\{
\begin{array}{ll}
\sum_{j\in N}\beta _{j}, & i=0, \\
v(\{0,i\})-\beta _{i}, & i\in N,%
\end{array}%
\right.
\end{equation*}%
where $\beta _{i}:=\min_{S\subseteq N,i\in S}\left\{
(v(S_{0})-v(S))/s\right\} $, $i\in N$. {$\varphi $ satisfies (EF), (SR) and
(RR) but not (PD).}
\end{example}

\section{Concluding remarks\label{SecConcl}}

In this paper we studied single-period distribution chains consisting of a
supplier and multiple non-competing retailers from a game-theoretic
viewpoint. All companies have incentives to cooperate since this results in
reduced costs and consequently in increased profits. Therefore, these chains
are analyzed by means of their corresponding cooperative games, the
RS-games. Among others it is shown that any RS-game has a nonempty core.
Further, any core-allocation has a natural interpretation in terms of its
underlying distribution chain. One such a core-allocation is the
mgpc-solution for RS-games, whose characterization is included. This
solution is fit for RS-games since it recognizes the importance of the
supplier in achieving full cooperation. These results imply that the
companies in a distribution chain are willing to cooperate because there
exist stable distributions of the joint profit, namely the core-allocations.
Further, the mgpc-solution is a suitable allocation since it is designed
especially for this kind of distribution chains.

Topics for further research are: (1) examine other
core-allocations like e.g. the nucleolus, and the $\tau $-value;
(2) investigate how the results change if the wholesale price
function is endogenous; (3) analyze what happens if the unit
production cost depends on the total quantity to be produced; (4)
study the new model and game that arise when considering the
warehouse as a decision-maker/player.


\vfill

\begin{thebibliography}{99}
\bibitem{BeFe03} F.\ Bernstein and A.\ Federgruen, Pricing and Replenishment
Strategies in a Distribution System with Competing Retailers, Operations
Research 51(3), 409--426 (2003).

\bibitem{BeFe05} F.\ Bernstein and A.\ Federgruen, Decentralized Supply
Chains with Competing Retailers under Demand Uncentainty, Management Science
51(1), 18--29 (2005).

\bibitem{B63} O.N.\ Bondareva, Some Applications of Linear Programming
Methods to the Theory of Cooperative Games (in Russian),\ Problemy Kibernety
10, 119--139 (1963).

\bibitem{BoHa01} P.\ Borm, H.\ Hamers and R.\ Hendrickx, Operations Research
Games: a Survey,\ TOP 9, 139--199 (2001).

\bibitem{CaLa01} G.P.\ Cachon and M.A.\ Lariviere, Turning the Supply Chain
into a Revenue Chain, Harvard Business Revieuw, 20--21 (2001).

\bibitem{CaLa05} G.P.\ Cachon and M.A.\ Lariviere, Supply Chain Coordination
with Revenue-Sharing Contracts: Strengths and Limitations, Management
Science 51, 30--44 (2005).

\bibitem{GuMe07} L.A.\ Guardiola, A.\ Meca and J.\ Puerto,
Production-Inventory Games: A New Class of Totally Balanced
Combinatorial Optimization Games, Forthcoming in Games and
Economic Behavior, (2007).

\bibitem{GuMe04b} L.A.\ Guardiola, A.\ Meca and J.\ Puerto,
On the core and Owen point of production-inventory games, CIO
paper I-2004-27, Miguel Hern\'{a}ndez University, Elche, Spain
(2004).

\bibitem{GuMe06} L.A.\ Guardiola, A.\ Meca and J.\ Puerto,
Coordination in periodic review inventory situations, CIO paper
I-2006-13, Miguel Hern\'{a}ndez University, Elche, Spain (2006).

\bibitem{HM89} S.\ Hart and A.\ Mas-Colell, Potential, Value, and
Consistency, Econometrica 57, 589--614 (1989).

\bibitem{KriKa01} H.\ Krishnan, R.\ Kapuscinski and D.\ Butz, Coordinating
Contracts for Decentralized Supply Chains with Retailer Promotional Effort,
Management Science 50(1), 48--63 (2001).

\bibitem{Mar05} E.\ Markakis and A.\ Saberi, On the core of the multicommodity flow game,
Decision Support Systems 39, 3--10 (2005).

\bibitem{Mi05} S.\ Minner, Bargaining for cooperative economic ordering,
Decision Support Systems (in press), 48--63 (2005).

\bibitem{OzFr04} U.\ \"{O}zen, J.\ Fransoo, H.W.\ Norde and M.\ Slikker,
Cooperation between Multiple Newsvendors with Warehouses, CentER DP 2004-34,
Tilburg University, Tilburg, The Netherlands (2004).

\bibitem{Robi93} L.W.\ Robinson, A Comment on Gerchak and Gupta's `On
Apportioning Costs to Customers in Centralized Continuous Review Inventory
Systems', Journal of Operations Management 11, 99--102 (1993).

\bibitem{SH53} L.S.\ Shapley, A Value for n-Person Games, in: H.\ Kuhn and
A.W.\ Tucker, Eds., Contributions to the Theory of Games II, 307--317
(Princeton University Press, Princeton, 1953).

\bibitem{SH67} L.S.\ Shapley, On Balanced Sets and Cores, Naval Research
Logististics 14, 453--460 (1967).

\bibitem{SH71} L.S.\ Shapley, Cores of Convex Games, International Journal
of Game Theory 1, 11--26 (1971).

\bibitem{SlFr01} M.\ Slikker, J.\ Fransoo and M.\ Wouters, Joint Ordering in
Multiple News-Vendor Situations: A Game Theoretical Approach, working paper,
Eindhoven University of Technology, Eindhoven, The Netherlands (2001).

\bibitem{SlFr05} M.\ Slikker, J.\ Fransoo and M.\ Wouters, Cooperation
between Multiple News-Vendors with Transshipments, European Journal of
Operational Research 167, 370--380 (2005).

\bibitem{Ta92} T.\ Taylor, Supply Chain Coordination under Channel Rebates
with Sales Effort Effects, Management Science 45, 1339--1385 (1992).

\bibitem{TsNa98} A.\ Tsay, S.\ Nahmias, N.\ Agrawal eds., Modeling Supply
Chain Contracts: A Review, Quantitave Models for Supply Chain Management,
Kluwer, Boston, MA (1998).

\bibitem{VdBo07} W.\ van den Heuvel, P.\ Borm and H.\ Hamers, Economic
Lot-sizing Games, forthcoming in European Journal of Operational
Research 176, 1117--1130 (2007).

\bibitem{RyMa00} G.\ van Ryzin and S.\ Mahajan, Supply Chain Coordination
under Horizontal Competition, Working paper, Columbia University, New York
(2000).

\bibitem{WaPa94} Q.\ Wang and M.\ Parlar, A Three-Person Game Theory Model
Arising in Stochastic Inventory Control, European Journal of Operational
Research 76, 83--97 (1994).
\end{thebibliography}
\end{document}